\documentclass[prl,
                reprint,
                amsmath,
                amssymb,
                showpacs,
                superscriptaddress]{revtex4-1}
\usepackage{graphicx} 
\usepackage{dcolumn}  
\usepackage{bm}       
\usepackage{amsfonts}
\usepackage{dsfont}
\usepackage{csquotes}

\usepackage{ulem}
\usepackage{color}

\newcommand{\Andre}{Andr{\'e  }}

\begin{document}


\title{High harmonic spectroscopy of disorder-induced Anderson localization}

\author{Adhip Pattanayak}
\affiliation{ Department of Physics, Indian Institute of Technology Bombay, Powai, Mumbai 400076, India }

\author{\'A. Jim\'enez-Gal\'an}
\affiliation{Max-Born Stra{\ss}e 2A, D-12489 Berlin, Germany}

\author{Misha Ivanov}
\affiliation{Max-Born Stra{\ss}e 2A, D-12489 Berlin, Germany}
\affiliation{Department of Physics, Humboldt University, Newtonstra{\ss}e 15, D-12489 Berlin, Germany.}
\affiliation{Blackett Laboratory, Imperial College London, South Kensington Campus, SW7 2AZ London, United Kingdom.}
                         
\author{Gopal Dixit}
\email[]{gdixit@phy.iitb.ac.in}
\affiliation{ Department of Physics, Indian Institute of Technology Bombay, Powai, Mumbai 400076, India }


\begin{abstract}
Exponential localization of wavefunctions in lattices, whether in real or synthetic dimensions, 
is a fundamental wave interference phenomenon. Localization of Bloch-type 
functions in space-periodic lattice, triggered
by spatial disorder, is known as Anderson localization and arrests 
diffusion of classical particles in disordered potentials. 
In time-periodic Floquet lattices, exponential localization
in a periodically driven quantum system similarly arrests diffusion of its classically chaotic counterpart
in the action-angle space.  Here we demonstrate that nonlinear optical response allows for 
clear detection of the disorder-induced phase transition between delocalized and localized states.
The optical signature of the transition is the emergence of symmetry-forbidden 
even-order harmonics: these harmonics are enabled by Anderson-type localization and 
arise for sufficiently strong disorder even when the overall charge distribution in the field-free system 
spatially symmetric. 
The ratio of even to odd harmonic intensities as a function of disorder 
maps out the phase transition even when the associated changes in the band structure 
are negligibly small.
\end{abstract}

\maketitle 

Disorder is an ubiquitous effect in crystals~\cite{RevModPhys.80.1355}. The seminal work by Anderson~\cite{Anderson} predicted that above a critical disorder value, the electronic 
wavefunction will change from being delocalized across the lattice to exponentially localized 
(insulating state) due to the interference of multiple quantum paths originating from the scattering with 
random impurities and defects. Anderson localization is a fundamental wave phenomenon 
and thus permeates many branches of physics; it has been observed in matter waves~\cite{Billy2008}, light waves~\cite{Wiersma1997}, and microwaves~\cite{Dalichaouch1991}. Anderson localization 
also finds direct analogues in periodically driven systems, with time-periodic 
dynamics taking the role of space-periodic structure. While 
periodically driven classical systems can develop chaotic behaviour for sufficiently strong
driving fields, leading to delocalization of the original ensemble across the whole phase
space, their quantum counterpart shows exponential localization of the light-dressed 
states~\cite{Casati1987}.

Dramatic changes in a wavefunction during transition from a delocalized to a localized
state may lead to changes in the nonlinear optical response of the system. 
In this context, symmetry--forbidden harmonics of the driving field are an appealing bellwether candidate.
Indeed, while even-order harmonics are known to be forbidden in systems with inversion symmetry~\cite{Boyd},
they are also known to arise in such systems if and when charges localize~\cite{SilvaIvanov2016,Bandrauk_2005}.
The required symmetry breaking can then be triggered by 
even a small asymmetry in the oscillating electric field of the 
driving laser pulse. Such asymmetry is natural in short laser pulses and is 
controlled by the phase of the electric field 
oscillations under the pulse envelope,  i.e., the carrier-envelope phase (CEP)~\cite{KrauszIvanovRevModPhys.81.163}.
Exponentially localized states in symmetric multiple well 
potentials appear to be particularly sensitive to even small 
field asymmetries, leading to even harmonics  in the nonlinear response even  
for pulses encompassing tens of cycles \cite{SilvaIvanov2016}.

High harmonic generation (HHG) is a powerful tool for ultrafast spectroscopy 
~\cite{smirnova2013multielectron, Lein_2007}.
Extremely large coherent bandwidth of harmonic spectra enables 
sub-femtosecond resolution. 
HHG is a sensitive probe of  
Cooper minima~\cite{VilleneuvePhysRevLett.109.143001}, Auger decay~\cite{LeeuwenburghPhysRevLett.111.123002}, attosecond dynamics of optical tunnelling~\cite{Shafir2012,Pedatzur2015} and the dynamics of
electron exchange~\cite{SukiasyanPhysRevLett.102.223002} in atoms,
ultrafast hole dynamics~\cite{Eckart2018,Gaal2007,BrunerReview,smirnova2013multielectron}, 
nuclear motion \cite{BaykushevaFD,Baker424,LeinPRL2017} in small molecules, and 
enantio-sensitive electronic response in more complex chiral molecules~\cite{Cireasa2015,Ayuso_2018,NeufeldPhysRevX,BaykushevaPNAS}. 
In solids, high harmonic spectroscopy has allowed observation of dynamical Bloch oscillations, band structure tomography~\cite{VampaPRL}, probing of defects in solids~\cite{AdhipPRA, mrudul2020high}, sub-fs monitoring of core excitons~\cite{LuuGoulielmakis2015}, optical measurement of the valley pseudospin~\cite{LangerHuber2018,Alvaro2020subcycle, jimenez2020light}, tracking of van Hove singularities~\cite{Uzan2020}, 
picometer resolution of valence band electrons~\cite{LakhotiaGoulielmakis2020}, 
imaging  internal structures of a unit cell~\cite{pattanayak2019direct}, monitoring of light-driven 
insulator-to-metal transitions~\cite{Silva2018}, and probing of topological effects~\cite{BauerPRLtopoedge,Silva2019}.

In this work, we employe HHG to track phase transition between delocalized and localized states
in the Aubry-\Andre (AA) system ~\cite{Aubry} (similar to that proposed in Ref.~\cite{Harper_1955}), where 
localization occurs only above a critical value of disorder, already in one dimension. 
This model captures the metal-to-insulator transition, is a workhorse to study non-trivial topology, and 
has been realized in optical lattices and photonic quasi-crystals~\cite{Sanchez-Palencia2010,Roati2008}.

The model system is described by the following tight-binding Hamiltonian,
\begin{equation}\label{eq:H_AA_without_field}
    \hat{H}  = -t_0\sum_{j=1}^{L-1}\left ( c_{j}^{\dagger}c_{j+1}+ \textrm{h.c.} \right ) 
     + V\sum_{j=1}^{L}\cos(2\pi\sigma j)c_{j}^{\dagger}c_{j},
\end{equation}
where $t_0$ is the nearest neighbour hopping term, $V$ is the strength of the potential, $c_j^\dagger$ and $c_j$ are, respectively, the fermionic creation and annihilation operators at site $j$, $ \textrm{h.c.}$ stands for the hermitian conjugate,  
$L$ is the total number of lattice sites and $\sigma$ determines the periodicity of the potential. 
A rational value of $\sigma$ corresponds to a periodic potential and consequently to delocalized electronic wavefunctions. If $\sigma$ is irrational, the potential becomes quasi-periodically disordered (for finite systems, this may also happen 
for rational $\sigma$).  
For a disordered potential,  the system undergoes 
the localization phase transition at $V/t_0 = 2$. 
For $V/t_0 > 2$ all states are exponentially localized on one site, while 
for $V/t_0 < 2$ the states are delocalized. 

\begin{figure}
    \centering
    \includegraphics[width=\linewidth]{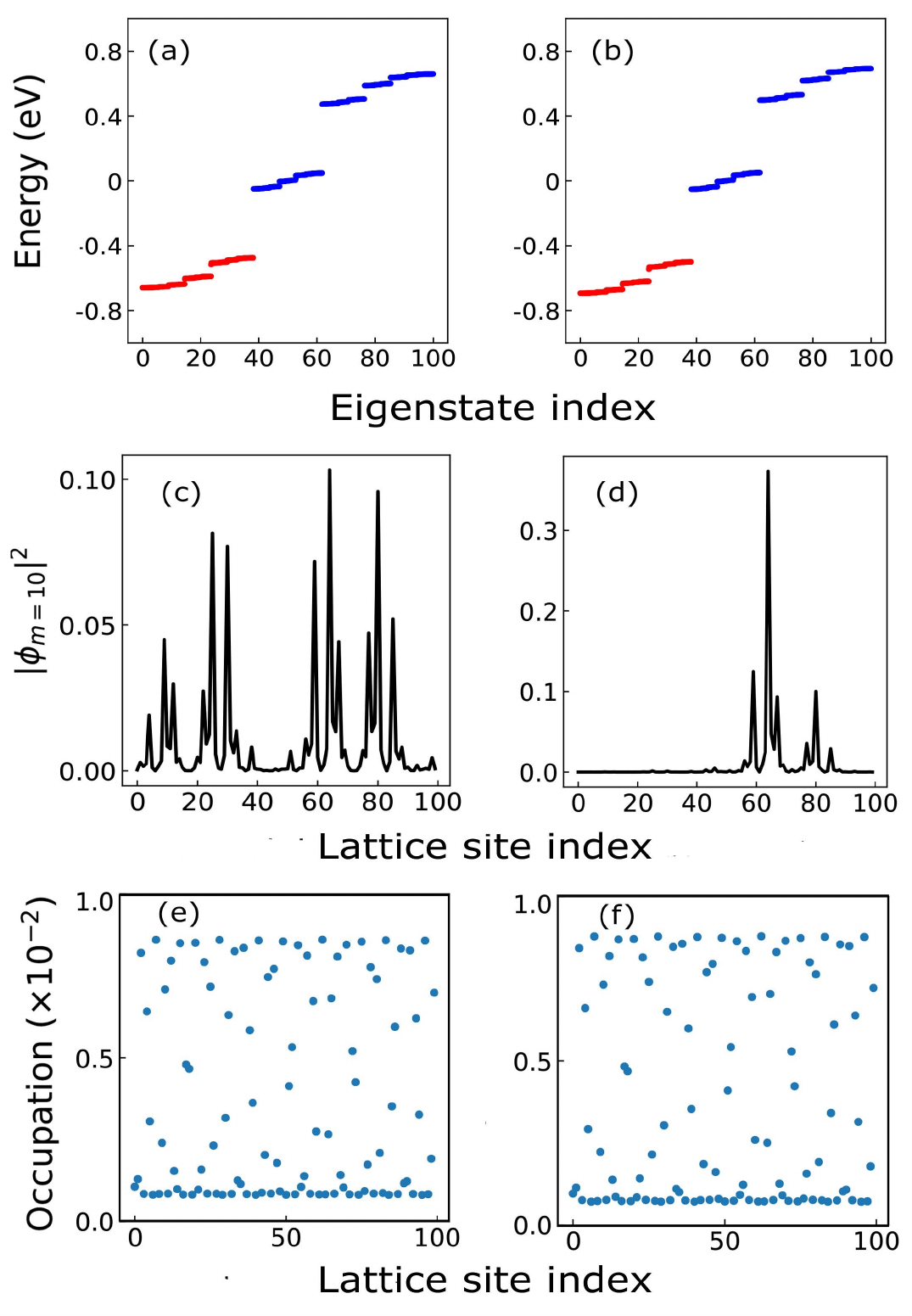}
    \caption{Electronic structure of the model system, for $\sigma=\frac{\sqrt{5}+1}{2}$. 
    Left panels (a,c,e) show the delocalized phase ($V/t_0=1.9$),  right panels (b,d,f) show 
    the localized phase ($V/t_0=2.1$). (a,b) -- eigenspectrum; (c,d) -- electron density per site for one eigenstate $(m=10)$, (e,f) lattice site occupation numbers in a fully-filled valence band. 
    Red and blue curves in (a) and (b) represent valence and conduction states, respectively.}
    \label{fig:fig1}
\end{figure}

Figures~\ref{fig:fig1}(a) and (b) show the eigenspectrum of our system, for 
$\sigma=\frac{\sqrt{5}+1}{2}$, with 100 lattice sites, in the delocalized ($V/t_0 = 1.9$) and localized ($V/t_0=2.1$) 
phases, respectively. 
The hopping term $t_0=0.26$ eV and the lattice constant $a_0=7.56$ atomic unit of length ($\sim$0.4 nm) are used throughout. The Fermi energy is $E_F=-0.2$~eV, so that red-colored states correspond to fully occupied valence band states, 
while the blue-colored states are unoccupied. 
Differences in the eigenspectrum for both phases are indiscernible. In contrast, the individual eigenstates of the system present a clear localization phase transition: Figs.~\ref{fig:fig1}(c) and (d)
show the occupation number of 
the eigenstate with index=10 (other eigenstates show similar behaviour). The eigenstate is delocalized 
for $V/t_0 = 1.9$ but fully localized for $V/t_0=2.1$. These differences between the charge distribution in 
individual eigenstates disappear completely when we consider the fully-filled valence 
band, see Figs.~\ref{fig:fig1}(e) and (f): the differences between both phases are barely visible. 

The question is: will the non-linear optical response be sensitive to the phase transition?
To address this question, we consider a non-resonant, low-frequency field polarized along the 1D chain,
\begin{equation}
F(t) = F_0 f(t) \cos (\omega t + \phi),
\end{equation}
with $\phi$ the carrier-envelope phase and $f(t)$ 
the sine-squared envelope with a full duration of 10 optical cycles. 
We include the laser-matter interaction via the time-dependent Peierls phase: 
$t_0 \to t_0\,e^{i a_0 e A(t)}$, 
where $A(t)$ is the field vector potential, $F(t)=-{dA(t)}/{dt}$, 
and $e$ is the electron charge.
The carrier $\omega=0.136$~eV is set well below the bandgap $\Delta\simeq 0.4$ eV.

Two characteristic regimes describe laser-induced electron dynamics in such low-frequency fields. 

In the localized phase, efficient resonant tunnelling
between localized states at different sites, including transitions from the valence to the conduction band,
becomes possible when the peak voltage between the adjacent sites $F_0a_0$ approaches and/or 
exceeds the characteristic energy gap $\Delta $, $F_0 a_0\simeq \Delta$. This regime enables rapid energy gain by 
the system within a few laser cycles, allowing it to climb  to the top of the energy scale~\cite{MishaPRA1996}.
In our case, its signature would be the emission of harmonics up to 
the maximum transition frequency of the system (harmonic 10), for
all fields enabling resonant tunnelling (RT). The onset of this regime corresponds 
to $F_{0,\text{RT}} \sim \Delta/a_0 \simeq  0.4$~MV/cm in our system.
In the localized phase, exponential sensitivity of resonant tunnelling to positions 
of individual states and to the field strength
create ideal conditions for symmetry breaking, leading to generation of even harmonics; the 
direction in which the symmetry is broken is controlled by the pulse CEP. 

In the delocalized case, the delocalized states adiabatically follow oscillations of the 
low-frequency driving field, preventing symmetry breaking. The latter requires 
light-induced electron localization, which occurs 
when  $F_0 a_0 \omega \geq \Delta^2$~\cite{DietrichIvanovPRL1996}, i.e., when   
$F_0 \geq 1 $  MV/cm in our system.
Even harmonic generation should therefore only emerge around $F_0 \simeq 1$ MV/cm.

Our numerical simulations below fully confirm all of these expectations. The time-dependent Schr\"odinger equation is solved independently for the $m$ normalized eigenstates $|\psi_m (t=t_i)\rangle$ of the field-free Hamiltonian Eq.~(\ref{eq:H_AA_without_field}) that lie below $E_F$, $| \psi_m (t) \rangle = e^{-i\int_{t_i}^{t}\hat{H}(t')dt'} |\psi_m (t=t_i)\rangle$, where $\hat{H}(t)$ is the Hamiltonian in Eq.~(\ref{eq:H_AA_without_field}) with the time-dependent Peierls substitution. The current from a single eigenstate $m$ is calculated as
\begin{equation}
j_m(t) = \left \langle \psi_m(t)|\hat{J}(t)|\psi_m(t) \right \rangle
\end{equation}
where the current operator is defined as
\begin{equation}
    \hat{J}(t) = -iea_0t_0\sum_{j=1}^{L}
    \left ( e^{-i a_0 e A(t)}c_{j}^{\dagger}c_{j+1}-e^{i a_0 e A(t)}c_{j+1}^{\dagger}c_{j} \right ).
\end{equation} 
The total current from all the valence band eigenstates $m$, i.e., the fully-filled valence band, is,
\begin{equation}
j (t) = \sum_m j_m (t).
\end{equation}
The harmonic spectra are then calculated from the Fourier transform of the time derivative of the 
total current. Prior to the Fourier transform, we apply an envelope to the current that coincides with the laser pulse envelope~\cite{MWuPRA}, to filter out emission after the end of the laser pulse. 
We consider 100 lattice sites and  0.01 atomic unit of time-step ($\sim$0.25 as) to get the converged spectra.

\begin{figure}
    \centering
    \includegraphics[width=\linewidth]{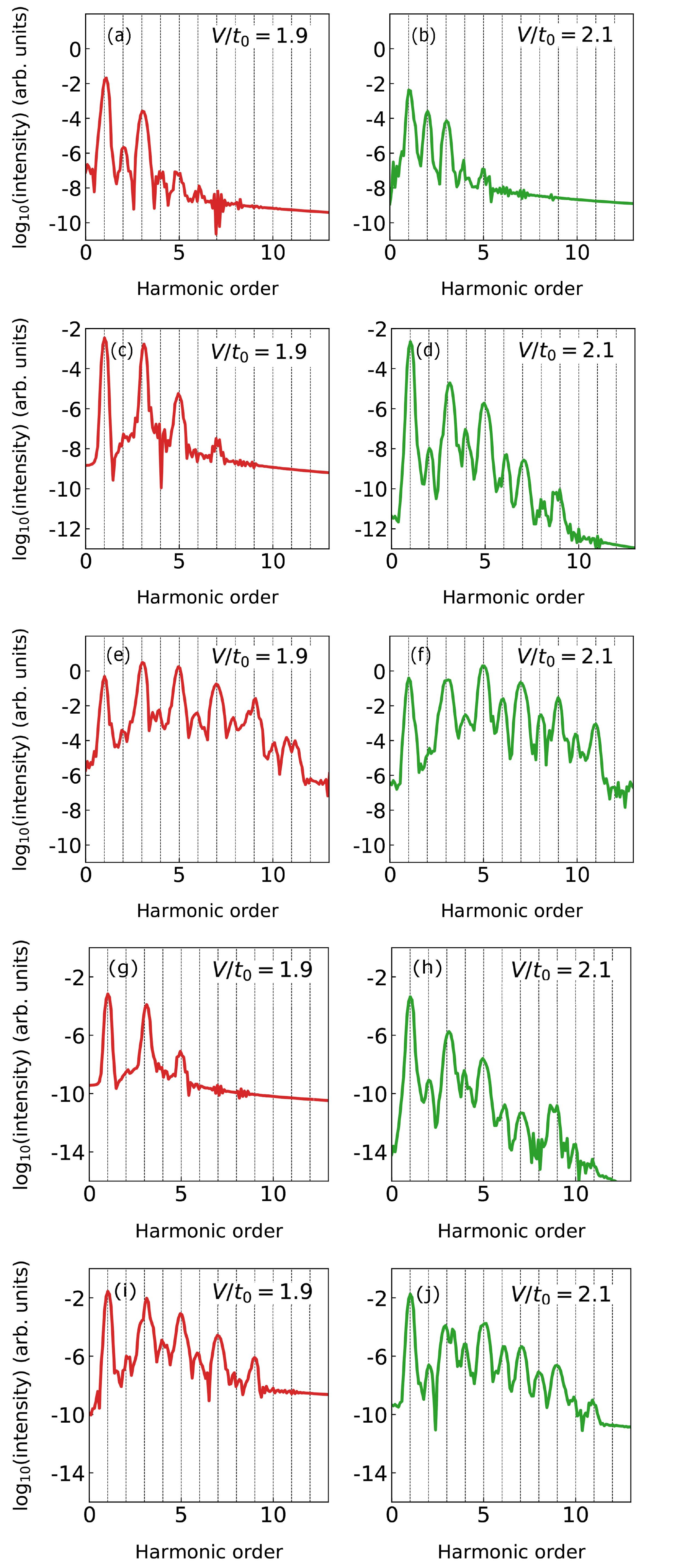}
    \caption{(a,b) High harmonic spectra for the system in the (a, c, e, g, i) delocalized and (b, d, f, h, j) localized phase: 
    (a, b) HHG from the 10$^{th}$ eigenstate only for a field strength $F_0 = 0.6$~MV/cm. 
  HHG from the fully-filled valence band for a field strength (c, d) $F_0 = 0.6$~MV/cm, (e, f)
  $F_0 = 2.2$~MV/cm, (e, f)  $F_0 = 0.4$~MV/cm, and (i, j)  $F_0 = 1.0$~MV/cm.}
    \label{fig:fig2}
\end{figure}

Figure~\ref{fig:fig2} shows the HHG spectrum for different initial states and field strengths in the localized and delocalized phases. First, in Figs.~\ref{fig:fig2}(a) and (b), we consider a single valence band eigenstate as our initial state ($m=10$, shown in Fig.~\ref{fig:fig1}(c,d)). The charge distribution of the initial state is strongly asymmetric in both phases, which breaks the left-right symmetry of the chain in the HHG process and leads to the appearance of even harmonics, both in the delocalized [Fig.\ref{fig:fig2}(a)] and localized [Fig.\ref{fig:fig2}(b)] phases.

\begin{figure}
    \centering
    \includegraphics[width=\linewidth]{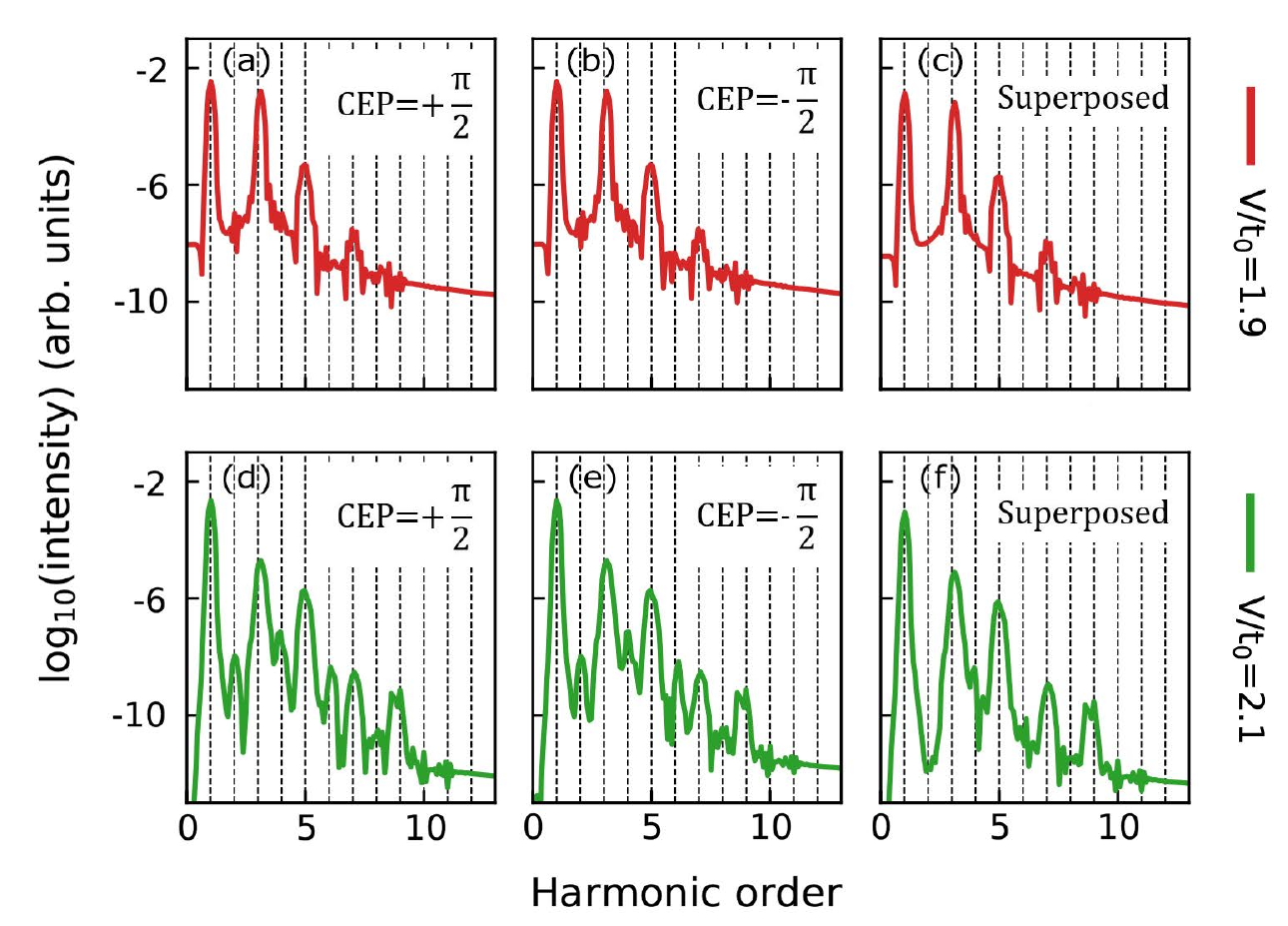}
    \caption{High harmonic spectra in the delocalized (top panels, red curve) and the localized (lower panels, green curve) phase calculated for different carrier-envelope phases (CEP) of the field: (a,d) CEP=$+\pi/2$, (b,e) CEP=$-\pi/2$ and (d,f) the coherent superposition of CEP=$+\pi/2$ and CEP=$-\pi/2$.}
    \label{fig:fig3}
\end{figure}

However, the equilibrium initial state corresponds to the fully-filled valence band, with 
the charge distributed relatively evenly between all sites; the distribution is virtually identical 
between the two phases [Figs.~\ref{fig:fig1}(e) and (f)]. 
With this initial condition, the HHG spectra of the 
two phases, shown in Figs.~\ref{fig:fig2}(c) and (d) for a field strength of $F_0 = 0.6$~MV/cm, are 
now strikingly different. 

Destructive interference from different initial states completely suppresses even harmonics in the 
delocalized phase [Fig.~\ref{fig:fig2}(c)], but
they remain prominent in the localized phase [Fig.~\ref{fig:fig2}(d)]. 
This result follows from our previous discussion. The field strength 
$F_0 = 0.6$~MV/cm is above the threshold for resonant tunneling between localized states in our system ($F_{0,\text{RT}} = 0.4$~MV/cm), where the instantaneous field brings the energy levels into resonance, 
generating coherence between all sites.  Resonant tunneling between the sites depends 
sensitively on the instantaneous field strength, leading to CEP-dependent 
symmetry breaking and the appearance of even harmonics in the localized phase. 
We find that even harmonics emerge already
at $F_0 \simeq 0.4$~MV/cm. As resonant tunnellng 
induces coherences between the localized sites, it enables population 
of the highest band and generation of higher harmonics in the 
localized phase than in the delocalized phase [Figs.~\ref{fig:fig2}(c) and (d)]. 
In the latter, the system follows adiabatically the field oscillations and 
transitions to the highest states are suppressed. 

The field strength used in Figs.~\ref{fig:fig2}(c) and (d) is below the threshold field for laser-induced localization
in the delocalized phase, which is $F_0 \sim 1$~MV/cm for our system. 
Therefore, the harmonic spectrum 
in the delocalized regime shows no sign of even harmonics and a smaller cut-off energy
[Fig.~\ref{fig:fig2}(c)]. Even harmonics  
emerge as soon as the field amplitude crosses this threshold. At  $F_0 = 2.2$~MV/cm 
the harmonic spectra in both phases become very similar [Figs.~\ref{fig:fig2}(e) and (f)]. 
For this field strength, the cut-off is the same in both phases, and corresponds to the 
(Stark-shifted) maximum transition frequency of the system.

To confirm the origin of symmetry breaking and even harmonics emission, Fig.~\ref{fig:fig3} shows the 
harmonic spectra for two values of the pulse carrier-envelope phase (CEP) shifted by $\pi$, 
and their coherent superposition in Figs.~\ref{fig:fig3}(c) and (f). 
In the delocalized case, even harmonics are absent regardless of the CEP 
[Figs.~\ref{fig:fig3}(a) and (b)]. 
In the localized 
case, they are identical in both cases, Figs.~\ref{fig:fig3}(d) and (e), 
but with opposite phase: upon coherent addition  
even harmonics are completely washed out [Fig.~\ref{fig:fig3}(f)]. 
The reason is that the laser-induced 
asymmetry in the electron charge distribution at CEP=$+\pi/2$ is exactly opposite to that at CEP=$-\pi/2$,
as graphically illustrated by the cancellation of the emission upon interference. 

\begin{figure}
    \centering
    \includegraphics[width=\linewidth]{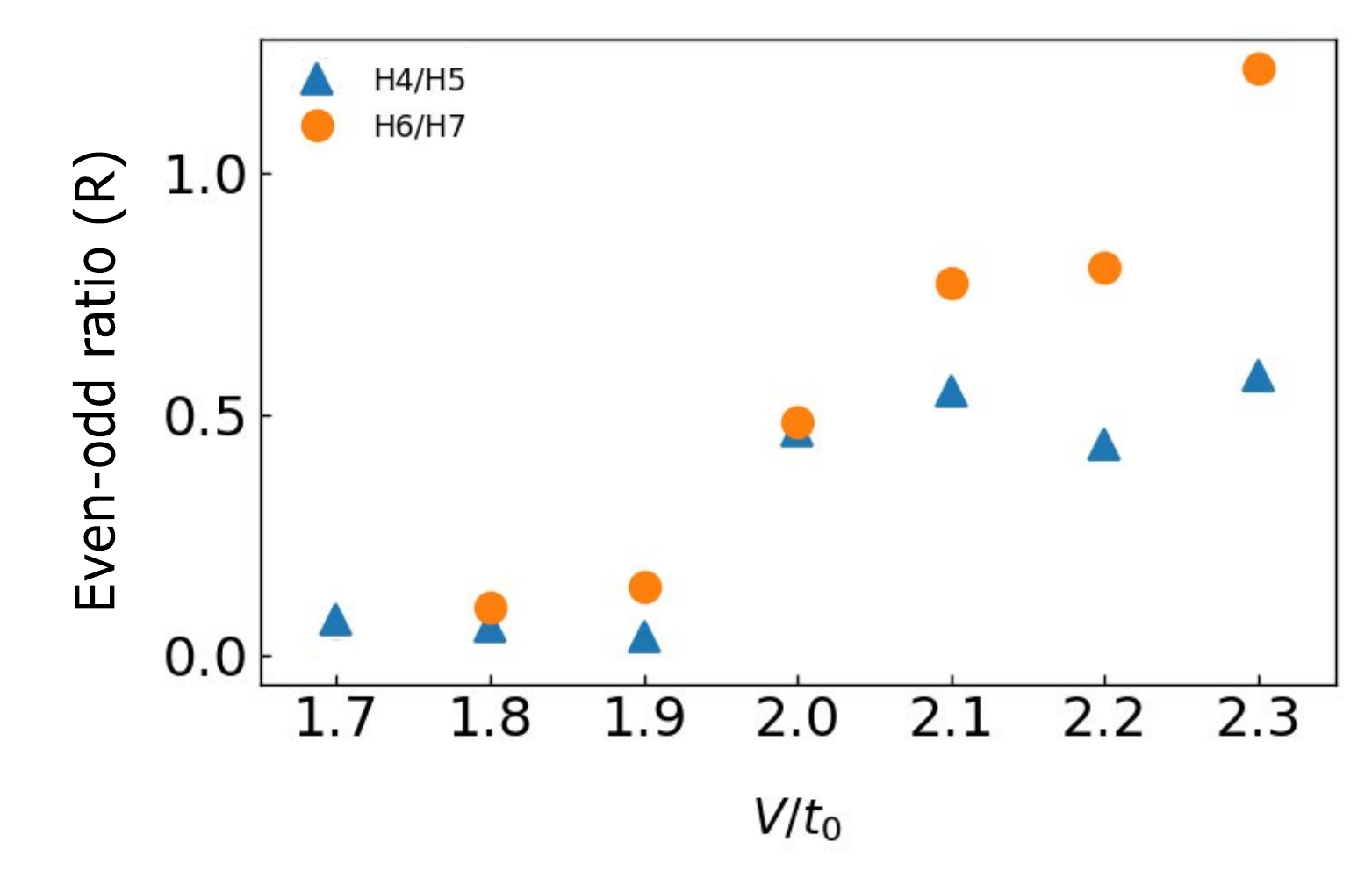}
    \caption{Even-to-odd harmonic peak ratio (R) for different $V/t_0$ ranging from 1.7 to 2.3. The ratio R is calculated as $\mathrm{R} = \mathrm{log_{10}} ( \bar{H}_{even} ) / \mathrm{log_{10}} ( \bar{H}_{odd} )$, where $\bar{H} = H^{peak} / H^{min}$. Blue tiangles (orange circles) represent the R value correspond to 4$^{th}$~(6$^{th}$) harmonic and 5$^{th}$~(7$^{th}$) harmonic. The increment of R value at $V/t_0>2$ indicates the phase transition phenomenon. }
    \label{fig:fig4}
\end{figure}

Figure~\ref{fig:fig4} shows that the relative intensity of even harmonics does indeed track the 
metal to insulator (delocalized -- localized) phase 
transition in the system. The Figure plots the ratio of even to odd harmonics for different 
values of $V/t_0$, at a field strength of $F_0 = 0.6$~MV/cm, i.e., that which generates 
resonant tunneling in the localized phase but not laser-induced localization in the delocalized phase. 
The even--odd ratio increases dramatically at the phase transition $V/t_0 = 2$, 
showing that HHG, driven by a phase-stable CEP pulse, is able to map disorder-induced 
electron localization in a solid.

In conclusion, we have used high harmonic spectroscopy to track delocalized to localized phase transition 
in the Aubry-Andr\'{e} system. For aperiodic potentials ($\sigma=\frac{\sqrt{5}+1}{2}$), the 
localized and delocalized phases show almost identical eigenspectra and site occupation numbers for a 
state with fully-filled valence band. Yet, high harmonic spectra between the two phases show 
striking differences, especially in the appearance of forbidden (even) harmonics. 
This effect is a consequence of dynamical symmetry breaking induced by the field but 
enabled by initial electron localization, which promotes resonant tunneling
and leads to CEP-dependent symmetry breaking even in the low-frequency regime. 
Our work shows that the localisation-delocalisation phase transition, which can be driven 
by a very small modification of the on-site energy, can be effectively traced by HHG spectroscopy.

A. P. acknowledges sandwich doctoral fellowship from Deutscher Akademischer Austauschdienst (DAAD, reference no. 57440919). 
M.I. and A. J-G acknowledge support of the FET-Open Optologic grant.  
M.I. acknowledges support from the Deutsche Forschungsgemeinschaft (DFG) Quantum Dynamics in Tailored Intense Fields (QUTIF) grant.
G. D. acknowledges financial support from Science and Engineering Research Board (SERB) India 
(Project No. ECR/2017/001460). 


\begin{thebibliography}{47}%
\makeatletter
\providecommand \@ifxundefined [1]{%
 \@ifx{#1\undefined}
}%
\providecommand \@ifnum [1]{%
 \ifnum #1\expandafter \@firstoftwo
 \else \expandafter \@secondoftwo
 \fi
}%
\providecommand \@ifx [1]{%
 \ifx #1\expandafter \@firstoftwo
 \else \expandafter \@secondoftwo
 \fi
}%
\providecommand \natexlab [1]{#1}%
\providecommand \enquote  [1]{``#1''}%
\providecommand \bibnamefont  [1]{#1}%
\providecommand \bibfnamefont [1]{#1}%
\providecommand \citenamefont [1]{#1}%
\providecommand \href@noop [0]{\@secondoftwo}%
\providecommand \href [0]{\begingroup \@sanitize@url \@href}%
\providecommand \@href[1]{\@@startlink{#1}\@@href}%
\providecommand \@@href[1]{\endgroup#1\@@endlink}%
\providecommand \@sanitize@url [0]{\catcode `\\12\catcode `\$12\catcode
  `\&12\catcode `\#12\catcode `\^12\catcode `\_12\catcode `\%12\relax}%
\providecommand \@@startlink[1]{}%
\providecommand \@@endlink[0]{}%
\providecommand \url  [0]{\begingroup\@sanitize@url \@url }%
\providecommand \@url [1]{\endgroup\@href {#1}{\urlprefix }}%
\providecommand \urlprefix  [0]{URL }%
\providecommand \Eprint [0]{\href }%
\providecommand \doibase [0]{http://dx.doi.org/}%
\providecommand \selectlanguage [0]{\@gobble}%
\providecommand \bibinfo  [0]{\@secondoftwo}%
\providecommand \bibfield  [0]{\@secondoftwo}%
\providecommand \translation [1]{[#1]}%
\providecommand \BibitemOpen [0]{}%
\providecommand \bibitemStop [0]{}%
\providecommand \bibitemNoStop [0]{.\EOS\space}%
\providecommand \EOS [0]{\spacefactor3000\relax}%
\providecommand \BibitemShut  [1]{\csname bibitem#1\endcsname}%
\let\auto@bib@innerbib\@empty
\bibitem [{\citenamefont {Evers}\ and\ \citenamefont
  {Mirlin}(2008)}]{RevModPhys.80.1355}%
  \BibitemOpen
  \bibfield  {author} {\bibinfo {author} {\bibfnamefont {F.}~\bibnamefont
  {Evers}}\ and\ \bibinfo {author} {\bibfnamefont {A.~D.}\ \bibnamefont
  {Mirlin}},\ }\href {\doibase 10.1103/RevModPhys.80.1355} {\bibfield
  {journal} {\bibinfo  {journal} {Rev. Mod. Phys.}\ }\textbf {\bibinfo {volume}
  {80}},\ \bibinfo {pages} {1355} (\bibinfo {year} {2008})}\BibitemShut
  {NoStop}%
\bibitem [{\citenamefont {Anderson}(1958)}]{Anderson}%
  \BibitemOpen
  \bibfield  {author} {\bibinfo {author} {\bibfnamefont {P.~W.}\ \bibnamefont
  {Anderson}},\ }\href {\doibase 10.1103/PhysRev.109.1492} {\bibfield
  {journal} {\bibinfo  {journal} {Phys. Rev.}\ }\textbf {\bibinfo {volume}
  {109}},\ \bibinfo {pages} {1492} (\bibinfo {year} {1958})}\BibitemShut
  {NoStop}%
\bibitem [{\citenamefont {Billy}\ \emph {et~al.}(2008)\citenamefont {Billy},
  \citenamefont {Josse}, \citenamefont {Zuo}, \citenamefont {Bernard},
  \citenamefont {Hambrecht}, \citenamefont {Lugan}, \citenamefont
  {Cl{\'e}ment}, \citenamefont {Sanchez-Palencia}, \citenamefont {Bouyer},\
  and\ \citenamefont {Aspect}}]{Billy2008}%
  \BibitemOpen
  \bibfield  {author} {\bibinfo {author} {\bibfnamefont {J.}~\bibnamefont
  {Billy}}, \bibinfo {author} {\bibfnamefont {V.}~\bibnamefont {Josse}},
  \bibinfo {author} {\bibfnamefont {Z.}~\bibnamefont {Zuo}}, \bibinfo {author}
  {\bibfnamefont {A.}~\bibnamefont {Bernard}}, \bibinfo {author} {\bibfnamefont
  {B.}~\bibnamefont {Hambrecht}}, \bibinfo {author} {\bibfnamefont
  {P.}~\bibnamefont {Lugan}}, \bibinfo {author} {\bibfnamefont
  {D.}~\bibnamefont {Cl{\'e}ment}}, \bibinfo {author} {\bibfnamefont
  {L.}~\bibnamefont {Sanchez-Palencia}}, \bibinfo {author} {\bibfnamefont
  {P.}~\bibnamefont {Bouyer}}, \ and\ \bibinfo {author} {\bibfnamefont
  {A.}~\bibnamefont {Aspect}},\ }\href {\doibase 10.1038/nature07000}
  {\bibfield  {journal} {\bibinfo  {journal} {Nature}\ }\textbf {\bibinfo
  {volume} {453}},\ \bibinfo {pages} {891} (\bibinfo {year}
  {2008})}\BibitemShut {NoStop}%
\bibitem [{\citenamefont {Wiersma}\ \emph {et~al.}(1997)\citenamefont
  {Wiersma}, \citenamefont {Bartolini}, \citenamefont {Lagendijk},\ and\
  \citenamefont {Righini}}]{Wiersma1997}%
  \BibitemOpen
  \bibfield  {author} {\bibinfo {author} {\bibfnamefont {D.~S.}\ \bibnamefont
  {Wiersma}}, \bibinfo {author} {\bibfnamefont {P.}~\bibnamefont {Bartolini}},
  \bibinfo {author} {\bibfnamefont {A.}~\bibnamefont {Lagendijk}}, \ and\
  \bibinfo {author} {\bibfnamefont {R.}~\bibnamefont {Righini}},\ }\href
  {\doibase 10.1038/37757} {\bibfield  {journal} {\bibinfo  {journal} {Nature}\
  }\textbf {\bibinfo {volume} {390}},\ \bibinfo {pages} {671} (\bibinfo {year}
  {1997})}\BibitemShut {NoStop}%
\bibitem [{\citenamefont {Dalichaouch}\ \emph {et~al.}(1991)\citenamefont
  {Dalichaouch}, \citenamefont {Armstrong}, \citenamefont {Schultz},
  \citenamefont {Platzman},\ and\ \citenamefont {McCall}}]{Dalichaouch1991}%
  \BibitemOpen
  \bibfield  {author} {\bibinfo {author} {\bibfnamefont {R.}~\bibnamefont
  {Dalichaouch}}, \bibinfo {author} {\bibfnamefont {J.~P.}\ \bibnamefont
  {Armstrong}}, \bibinfo {author} {\bibfnamefont {S.}~\bibnamefont {Schultz}},
  \bibinfo {author} {\bibfnamefont {P.~M.}\ \bibnamefont {Platzman}}, \ and\
  \bibinfo {author} {\bibfnamefont {S.~L.}\ \bibnamefont {McCall}},\ }\href
  {\doibase 10.1038/354053a0} {\bibfield  {journal} {\bibinfo  {journal}
  {Nature}\ }\textbf {\bibinfo {volume} {354}},\ \bibinfo {pages} {53}
  (\bibinfo {year} {1991})}\BibitemShut {NoStop}%
\bibitem [{\citenamefont {Casati}\ \emph {et~al.}(1987)\citenamefont {Casati},
  \citenamefont {Chirikov}, \citenamefont {Shepelyansky},\ and\ \citenamefont
  {Guarneri}}]{Casati1987}%
  \BibitemOpen
  \bibfield  {author} {\bibinfo {author} {\bibfnamefont {G.}~\bibnamefont
  {Casati}}, \bibinfo {author} {\bibfnamefont {B.~V.}\ \bibnamefont
  {Chirikov}}, \bibinfo {author} {\bibfnamefont {D.~L.}\ \bibnamefont
  {Shepelyansky}}, \ and\ \bibinfo {author} {\bibfnamefont {I.}~\bibnamefont
  {Guarneri}},\ }\href
  {http://www.sciencedirect.com/science/article/pii/0370157387900093}
  {\bibfield  {journal} {\bibinfo  {journal} {Physics Reports}\ }\textbf
  {\bibinfo {volume} {154}},\ \bibinfo {pages} {77} (\bibinfo {year}
  {1987})}\BibitemShut {NoStop}%
\bibitem [{\citenamefont {Boyd}(2008)}]{Boyd}%
  \BibitemOpen
  \bibfield  {author} {\bibinfo {author} {\bibfnamefont {R.~W.}\ \bibnamefont
  {Boyd}},\ }\href@noop {} {\enquote {\bibinfo {title} {Nonlinear optics, third
  edition},}\ } (\bibinfo {year} {2008})\BibitemShut {NoStop}%
\bibitem [{\citenamefont {Silva}\ \emph {et~al.}(2016)\citenamefont {Silva},
  \citenamefont {Rivi{\'e}re}, \citenamefont {Morales}, \citenamefont
  {Smirnova}, \citenamefont {Ivanov},\ and\ \citenamefont
  {Mart{\'i}n}}]{SilvaIvanov2016}%
  \BibitemOpen
  \bibfield  {author} {\bibinfo {author} {\bibfnamefont {R.~E.~F.}\
  \bibnamefont {Silva}}, \bibinfo {author} {\bibfnamefont {P.}~\bibnamefont
  {Rivi{\'e}re}}, \bibinfo {author} {\bibfnamefont {F.}~\bibnamefont
  {Morales}}, \bibinfo {author} {\bibfnamefont {O.}~\bibnamefont {Smirnova}},
  \bibinfo {author} {\bibfnamefont {M.}~\bibnamefont {Ivanov}}, \ and\ \bibinfo
  {author} {\bibfnamefont {F.}~\bibnamefont {Mart{\'i}n}},\ }\href {\doibase
  10.1038/srep32653} {\bibfield  {journal} {\bibinfo  {journal} {Scientific
  Reports}\ }\textbf {\bibinfo {volume} {6}},\ \bibinfo {pages} {32653}
  (\bibinfo {year} {2016})}\BibitemShut {NoStop}%
\bibitem [{\citenamefont {Bandrauk}\ and\ \citenamefont
  {Lu}(2005)}]{Bandrauk_2005}%
  \BibitemOpen
  \bibfield  {author} {\bibinfo {author} {\bibfnamefont {A.~D.}\ \bibnamefont
  {Bandrauk}}\ and\ \bibinfo {author} {\bibfnamefont {H.}~\bibnamefont {Lu}},\
  }\href {\doibase 10.1088/0953-4075/38/14/016} {\bibfield  {journal} {\bibinfo
   {journal} {Journal of Physics B: Atomic, Molecular and Optical Physics}\
  }\textbf {\bibinfo {volume} {38}},\ \bibinfo {pages} {2529} (\bibinfo {year}
  {2005})}\BibitemShut {NoStop}%
\bibitem [{\citenamefont {Krausz}\ and\ \citenamefont
  {Ivanov}(2009)}]{KrauszIvanovRevModPhys.81.163}%
  \BibitemOpen
  \bibfield  {author} {\bibinfo {author} {\bibfnamefont {F.}~\bibnamefont
  {Krausz}}\ and\ \bibinfo {author} {\bibfnamefont {M.}~\bibnamefont
  {Ivanov}},\ }\href {\doibase 10.1103/RevModPhys.81.163} {\bibfield  {journal}
  {\bibinfo  {journal} {Rev. Mod. Phys.}\ }\textbf {\bibinfo {volume} {81}},\
  \bibinfo {pages} {163} (\bibinfo {year} {2009})}\BibitemShut {NoStop}%
\bibitem [{\citenamefont {Smirnova}\ and\ \citenamefont
  {Ivanov}(2013)}]{smirnova2013multielectron}%
  \BibitemOpen
  \bibfield  {author} {\bibinfo {author} {\bibfnamefont {O.}~\bibnamefont
  {Smirnova}}\ and\ \bibinfo {author} {\bibfnamefont {M.}~\bibnamefont
  {Ivanov}},\ }\href@noop {} {\enquote {\bibinfo {title} {Multielectron high
  harmonic generation: simple man on a complex plane},}\ } (\bibinfo {year}
  {2013}),\ \Eprint {http://arxiv.org/abs/1304.2413} {arXiv:1304.2413
  [physics.atom-ph]} \BibitemShut {NoStop}%
\bibitem [{\citenamefont {Lein}(2007)}]{Lein_2007}%
  \BibitemOpen
  \bibfield  {author} {\bibinfo {author} {\bibfnamefont {M.}~\bibnamefont
  {Lein}},\ }\href {\doibase 10.1088/0953-4075/40/16/r01} {\bibfield  {journal}
  {\bibinfo  {journal} {Journal of Physics B: Atomic, Molecular and Optical
  Physics}\ }\textbf {\bibinfo {volume} {40}},\ \bibinfo {pages} {R135}
  (\bibinfo {year} {2007})}\BibitemShut {NoStop}%
\bibitem [{\citenamefont {Bertrand}\ \emph {et~al.}(2012)\citenamefont
  {Bertrand}, \citenamefont {W\"orner}, \citenamefont {Hockett}, \citenamefont
  {Villeneuve},\ and\ \citenamefont
  {Corkum}}]{VilleneuvePhysRevLett.109.143001}%
  \BibitemOpen
  \bibfield  {author} {\bibinfo {author} {\bibfnamefont {J.~B.}\ \bibnamefont
  {Bertrand}}, \bibinfo {author} {\bibfnamefont {H.~J.}\ \bibnamefont
  {W\"orner}}, \bibinfo {author} {\bibfnamefont {P.}~\bibnamefont {Hockett}},
  \bibinfo {author} {\bibfnamefont {D.~M.}\ \bibnamefont {Villeneuve}}, \ and\
  \bibinfo {author} {\bibfnamefont {P.~B.}\ \bibnamefont {Corkum}},\ }\href
  {\doibase 10.1103/PhysRevLett.109.143001} {\bibfield  {journal} {\bibinfo
  {journal} {Phys. Rev. Lett.}\ }\textbf {\bibinfo {volume} {109}},\ \bibinfo
  {pages} {143001} (\bibinfo {year} {2012})}\BibitemShut {NoStop}%
\bibitem [{\citenamefont {Leeuwenburgh}\ \emph {et~al.}(2013)\citenamefont
  {Leeuwenburgh}, \citenamefont {Cooper}, \citenamefont {Averbukh},
  \citenamefont {Marangos},\ and\ \citenamefont
  {Ivanov}}]{LeeuwenburghPhysRevLett.111.123002}%
  \BibitemOpen
  \bibfield  {author} {\bibinfo {author} {\bibfnamefont {J.}~\bibnamefont
  {Leeuwenburgh}}, \bibinfo {author} {\bibfnamefont {B.}~\bibnamefont
  {Cooper}}, \bibinfo {author} {\bibfnamefont {V.}~\bibnamefont {Averbukh}},
  \bibinfo {author} {\bibfnamefont {J.~P.}\ \bibnamefont {Marangos}}, \ and\
  \bibinfo {author} {\bibfnamefont {M.}~\bibnamefont {Ivanov}},\ }\href
  {\doibase 10.1103/PhysRevLett.111.123002} {\bibfield  {journal} {\bibinfo
  {journal} {Phys. Rev. Lett.}\ }\textbf {\bibinfo {volume} {111}},\ \bibinfo
  {pages} {123002} (\bibinfo {year} {2013})}\BibitemShut {NoStop}%
\bibitem [{\citenamefont {Shafir}\ \emph {et~al.}(2012)\citenamefont {Shafir},
  \citenamefont {Soifer}, \citenamefont {Bruner}, \citenamefont {Dagan},
  \citenamefont {Mairesse}, \citenamefont {Patchkovskii}, \citenamefont
  {Ivanov}, \citenamefont {Smirnova},\ and\ \citenamefont
  {Dudovich}}]{Shafir2012}%
  \BibitemOpen
  \bibfield  {author} {\bibinfo {author} {\bibfnamefont {D.}~\bibnamefont
  {Shafir}}, \bibinfo {author} {\bibfnamefont {H.}~\bibnamefont {Soifer}},
  \bibinfo {author} {\bibfnamefont {B.~D.}\ \bibnamefont {Bruner}}, \bibinfo
  {author} {\bibfnamefont {M.}~\bibnamefont {Dagan}}, \bibinfo {author}
  {\bibfnamefont {Y.}~\bibnamefont {Mairesse}}, \bibinfo {author}
  {\bibfnamefont {S.}~\bibnamefont {Patchkovskii}}, \bibinfo {author}
  {\bibfnamefont {M.~Y.}\ \bibnamefont {Ivanov}}, \bibinfo {author}
  {\bibfnamefont {O.}~\bibnamefont {Smirnova}}, \ and\ \bibinfo {author}
  {\bibfnamefont {N.}~\bibnamefont {Dudovich}},\ }\href {\doibase
  10.1038/nature11025} {\bibfield  {journal} {\bibinfo  {journal} {Nature}\
  }\textbf {\bibinfo {volume} {485}},\ \bibinfo {pages} {343} (\bibinfo {year}
  {2012})}\BibitemShut {NoStop}%
\bibitem [{\citenamefont {Pedatzur}\ \emph {et~al.}(2015)\citenamefont
  {Pedatzur}, \citenamefont {Orenstein}, \citenamefont {Serbinenko},
  \citenamefont {Soifer}, \citenamefont {Bruner}, \citenamefont {Uzan},
  \citenamefont {Brambila}, \citenamefont {Harvey}, \citenamefont {Torlina},
  \citenamefont {Morales}, \citenamefont {Smirnova},\ and\ \citenamefont
  {Dudovich}}]{Pedatzur2015}%
  \BibitemOpen
  \bibfield  {author} {\bibinfo {author} {\bibfnamefont {O.}~\bibnamefont
  {Pedatzur}}, \bibinfo {author} {\bibfnamefont {G.}~\bibnamefont {Orenstein}},
  \bibinfo {author} {\bibfnamefont {V.}~\bibnamefont {Serbinenko}}, \bibinfo
  {author} {\bibfnamefont {H.}~\bibnamefont {Soifer}}, \bibinfo {author}
  {\bibfnamefont {B.~D.}\ \bibnamefont {Bruner}}, \bibinfo {author}
  {\bibfnamefont {A.~J.}\ \bibnamefont {Uzan}}, \bibinfo {author}
  {\bibfnamefont {D.~S.}\ \bibnamefont {Brambila}}, \bibinfo {author}
  {\bibfnamefont {A.~{\^A}.~G.}\ \bibnamefont {Harvey}}, \bibinfo {author}
  {\bibfnamefont {L.}~\bibnamefont {Torlina}}, \bibinfo {author} {\bibfnamefont
  {F.}~\bibnamefont {Morales}}, \bibinfo {author} {\bibfnamefont
  {O.}~\bibnamefont {Smirnova}}, \ and\ \bibinfo {author} {\bibfnamefont
  {N.}~\bibnamefont {Dudovich}},\ }\href {\doibase 10.1038/nphys3436}
  {\bibfield  {journal} {\bibinfo  {journal} {Nature Physics}\ }\textbf
  {\bibinfo {volume} {11}},\ \bibinfo {pages} {815} (\bibinfo {year}
  {2015})}\BibitemShut {NoStop}%
\bibitem [{\citenamefont {Sukiasyan}\ \emph {et~al.}(2009)\citenamefont
  {Sukiasyan}, \citenamefont {McDonald}, \citenamefont {Destefani},
  \citenamefont {Ivanov},\ and\ \citenamefont
  {Brabec}}]{SukiasyanPhysRevLett.102.223002}%
  \BibitemOpen
  \bibfield  {author} {\bibinfo {author} {\bibfnamefont {S.}~\bibnamefont
  {Sukiasyan}}, \bibinfo {author} {\bibfnamefont {C.}~\bibnamefont {McDonald}},
  \bibinfo {author} {\bibfnamefont {C.}~\bibnamefont {Destefani}}, \bibinfo
  {author} {\bibfnamefont {M.~Y.}\ \bibnamefont {Ivanov}}, \ and\ \bibinfo
  {author} {\bibfnamefont {T.}~\bibnamefont {Brabec}},\ }\href {\doibase
  10.1103/PhysRevLett.102.223002} {\bibfield  {journal} {\bibinfo  {journal}
  {Phys. Rev. Lett.}\ }\textbf {\bibinfo {volume} {102}},\ \bibinfo {pages}
  {223002} (\bibinfo {year} {2009})}\BibitemShut {NoStop}%
\bibitem [{\citenamefont {Eckart}\ \emph {et~al.}(2018)\citenamefont {Eckart},
  \citenamefont {Kunitski}, \citenamefont {Richter}, \citenamefont {Hartung},
  \citenamefont {Rist}, \citenamefont {Trinter}, \citenamefont {Fehre},
  \citenamefont {Schlott}, \citenamefont {Henrichs}, \citenamefont {Schmidt},
  \citenamefont {Jahnke}, \citenamefont {Sch{\"o}ffler}, \citenamefont {Liu},
  \citenamefont {Barth}, \citenamefont {Kaushal}, \citenamefont {Morales},
  \citenamefont {Ivanov}, \citenamefont {Smirnova},\ and\ \citenamefont
  {D{\"o}rner}}]{Eckart2018}%
  \BibitemOpen
  \bibfield  {author} {\bibinfo {author} {\bibfnamefont {S.}~\bibnamefont
  {Eckart}}, \bibinfo {author} {\bibfnamefont {M.}~\bibnamefont {Kunitski}},
  \bibinfo {author} {\bibfnamefont {M.}~\bibnamefont {Richter}}, \bibinfo
  {author} {\bibfnamefont {A.}~\bibnamefont {Hartung}}, \bibinfo {author}
  {\bibfnamefont {J.}~\bibnamefont {Rist}}, \bibinfo {author} {\bibfnamefont
  {F.}~\bibnamefont {Trinter}}, \bibinfo {author} {\bibfnamefont
  {K.}~\bibnamefont {Fehre}}, \bibinfo {author} {\bibfnamefont
  {N.}~\bibnamefont {Schlott}}, \bibinfo {author} {\bibfnamefont
  {K.}~\bibnamefont {Henrichs}}, \bibinfo {author} {\bibfnamefont {L.~P.~H.}\
  \bibnamefont {Schmidt}}, \bibinfo {author} {\bibfnamefont {T.}~\bibnamefont
  {Jahnke}}, \bibinfo {author} {\bibfnamefont {M.}~\bibnamefont
  {Sch{\"o}ffler}}, \bibinfo {author} {\bibfnamefont {K.}~\bibnamefont {Liu}},
  \bibinfo {author} {\bibfnamefont {I.}~\bibnamefont {Barth}}, \bibinfo
  {author} {\bibfnamefont {J.}~\bibnamefont {Kaushal}}, \bibinfo {author}
  {\bibfnamefont {F.}~\bibnamefont {Morales}}, \bibinfo {author} {\bibfnamefont
  {M.}~\bibnamefont {Ivanov}}, \bibinfo {author} {\bibfnamefont
  {O.}~\bibnamefont {Smirnova}}, \ and\ \bibinfo {author} {\bibfnamefont
  {R.}~\bibnamefont {D{\"o}rner}},\ }\href {\doibase 10.1038/s41567-018-0080-5}
  {\bibfield  {journal} {\bibinfo  {journal} {Nature Physics}\ }\textbf
  {\bibinfo {volume} {14}},\ \bibinfo {pages} {701} (\bibinfo {year}
  {2018})}\BibitemShut {NoStop}%
\bibitem [{\citenamefont {Gaal}\ \emph {et~al.}(2007)\citenamefont {Gaal},
  \citenamefont {Kuehn}, \citenamefont {Reimann}, \citenamefont {Woerner},
  \citenamefont {Elsaesser},\ and\ \citenamefont {Hey}}]{Gaal2007}%
  \BibitemOpen
  \bibfield  {author} {\bibinfo {author} {\bibfnamefont {P.}~\bibnamefont
  {Gaal}}, \bibinfo {author} {\bibfnamefont {W.}~\bibnamefont {Kuehn}},
  \bibinfo {author} {\bibfnamefont {K.}~\bibnamefont {Reimann}}, \bibinfo
  {author} {\bibfnamefont {M.}~\bibnamefont {Woerner}}, \bibinfo {author}
  {\bibfnamefont {T.}~\bibnamefont {Elsaesser}}, \ and\ \bibinfo {author}
  {\bibfnamefont {R.}~\bibnamefont {Hey}},\ }\href {\doibase
  10.1038/nature06399} {\bibfield  {journal} {\bibinfo  {journal} {Nature}\
  }\textbf {\bibinfo {volume} {450}},\ \bibinfo {pages} {1210} (\bibinfo {year}
  {2007})}\BibitemShut {NoStop}%
\bibitem [{\citenamefont {Bruner}\ \emph {et~al.}(2016)\citenamefont {Bruner},
  \citenamefont {Mašín}, \citenamefont {Negro}, \citenamefont {Morales},
  \citenamefont {Brambila}, \citenamefont {Devetta}, \citenamefont {Faccialà},
  \citenamefont {Harvey}, \citenamefont {Ivanov}, \citenamefont {Mairesse},
  \citenamefont {Patchkovskii}, \citenamefont {Serbinenko}, \citenamefont
  {Soifer}, \citenamefont {Stagira}, \citenamefont {Vozzi}, \citenamefont
  {Dudovich},\ and\ \citenamefont {Smirnova}}]{BrunerReview}%
  \BibitemOpen
  \bibfield  {author} {\bibinfo {author} {\bibfnamefont {B.~D.}\ \bibnamefont
  {Bruner}}, \bibinfo {author} {\bibfnamefont {Z.}~\bibnamefont {Mašín}},
  \bibinfo {author} {\bibfnamefont {M.}~\bibnamefont {Negro}}, \bibinfo
  {author} {\bibfnamefont {F.}~\bibnamefont {Morales}}, \bibinfo {author}
  {\bibfnamefont {D.}~\bibnamefont {Brambila}}, \bibinfo {author}
  {\bibfnamefont {M.}~\bibnamefont {Devetta}}, \bibinfo {author} {\bibfnamefont
  {D.}~\bibnamefont {Faccialà}}, \bibinfo {author} {\bibfnamefont {A.~G.}\
  \bibnamefont {Harvey}}, \bibinfo {author} {\bibfnamefont {M.}~\bibnamefont
  {Ivanov}}, \bibinfo {author} {\bibfnamefont {Y.}~\bibnamefont {Mairesse}},
  \bibinfo {author} {\bibfnamefont {S.}~\bibnamefont {Patchkovskii}}, \bibinfo
  {author} {\bibfnamefont {V.}~\bibnamefont {Serbinenko}}, \bibinfo {author}
  {\bibfnamefont {H.}~\bibnamefont {Soifer}}, \bibinfo {author} {\bibfnamefont
  {S.}~\bibnamefont {Stagira}}, \bibinfo {author} {\bibfnamefont
  {C.}~\bibnamefont {Vozzi}}, \bibinfo {author} {\bibfnamefont
  {N.}~\bibnamefont {Dudovich}}, \ and\ \bibinfo {author} {\bibfnamefont
  {O.}~\bibnamefont {Smirnova}},\ }\href {\doibase 10.1039/C6FD00130K}
  {\bibfield  {journal} {\bibinfo  {journal} {Faraday Discuss.}\ }\textbf
  {\bibinfo {volume} {194}},\ \bibinfo {pages} {369} (\bibinfo {year}
  {2016})}\BibitemShut {NoStop}%
\bibitem [{\citenamefont {Baykusheva}\ \emph {et~al.}(2014)\citenamefont
  {Baykusheva}, \citenamefont {Kraus}, \citenamefont {Zhang}, \citenamefont
  {Rohringer},\ and\ \citenamefont {W{\"o}rner}}]{BaykushevaFD}%
  \BibitemOpen
  \bibfield  {author} {\bibinfo {author} {\bibfnamefont {D.}~\bibnamefont
  {Baykusheva}}, \bibinfo {author} {\bibfnamefont {P.~M.}\ \bibnamefont
  {Kraus}}, \bibinfo {author} {\bibfnamefont {S.~B.}\ \bibnamefont {Zhang}},
  \bibinfo {author} {\bibfnamefont {N.}~\bibnamefont {Rohringer}}, \ and\
  \bibinfo {author} {\bibfnamefont {H.~J.}\ \bibnamefont {W{\"o}rner}},\ }\href
  {\doibase 10.1039/C4FD00018H} {\bibfield  {journal} {\bibinfo  {journal}
  {Faraday Discuss.}\ }\textbf {\bibinfo {volume} {171}},\ \bibinfo {pages}
  {113} (\bibinfo {year} {2014})}\BibitemShut {NoStop}%
\bibitem [{\citenamefont {Baker}\ \emph {et~al.}(2006)\citenamefont {Baker},
  \citenamefont {Robinson}, \citenamefont {Haworth}, \citenamefont {Teng},
  \citenamefont {Smith}, \citenamefont {Chiril{\u a}}, \citenamefont {Lein},
  \citenamefont {Tisch},\ and\ \citenamefont {Marangos}}]{Baker424}%
  \BibitemOpen
  \bibfield  {author} {\bibinfo {author} {\bibfnamefont {S.}~\bibnamefont
  {Baker}}, \bibinfo {author} {\bibfnamefont {J.~S.}\ \bibnamefont {Robinson}},
  \bibinfo {author} {\bibfnamefont {C.~A.}\ \bibnamefont {Haworth}}, \bibinfo
  {author} {\bibfnamefont {H.}~\bibnamefont {Teng}}, \bibinfo {author}
  {\bibfnamefont {R.~A.}\ \bibnamefont {Smith}}, \bibinfo {author}
  {\bibfnamefont {C.~C.}\ \bibnamefont {Chiril{\u a}}}, \bibinfo {author}
  {\bibfnamefont {M.}~\bibnamefont {Lein}}, \bibinfo {author} {\bibfnamefont
  {J.~W.~G.}\ \bibnamefont {Tisch}}, \ and\ \bibinfo {author} {\bibfnamefont
  {J.~P.}\ \bibnamefont {Marangos}},\ }\href {\doibase 10.1126/science.1123904}
  {\bibfield  {journal} {\bibinfo  {journal} {Science}\ }\textbf {\bibinfo
  {volume} {312}},\ \bibinfo {pages} {424} (\bibinfo {year}
  {2006})}\BibitemShut {NoStop}%
\bibitem [{\citenamefont {Lan}\ \emph {et~al.}(2017)\citenamefont {Lan},
  \citenamefont {Ruhmann}, \citenamefont {He}, \citenamefont {Zhai},
  \citenamefont {Wang}, \citenamefont {Zhu}, \citenamefont {Zhang},
  \citenamefont {Zhou}, \citenamefont {Li}, \citenamefont {Lein},\ and\
  \citenamefont {Lu}}]{LeinPRL2017}%
  \BibitemOpen
  \bibfield  {author} {\bibinfo {author} {\bibfnamefont {P.}~\bibnamefont
  {Lan}}, \bibinfo {author} {\bibfnamefont {M.}~\bibnamefont {Ruhmann}},
  \bibinfo {author} {\bibfnamefont {L.}~\bibnamefont {He}}, \bibinfo {author}
  {\bibfnamefont {C.}~\bibnamefont {Zhai}}, \bibinfo {author} {\bibfnamefont
  {F.}~\bibnamefont {Wang}}, \bibinfo {author} {\bibfnamefont {X.}~\bibnamefont
  {Zhu}}, \bibinfo {author} {\bibfnamefont {Q.}~\bibnamefont {Zhang}}, \bibinfo
  {author} {\bibfnamefont {Y.}~\bibnamefont {Zhou}}, \bibinfo {author}
  {\bibfnamefont {M.}~\bibnamefont {Li}}, \bibinfo {author} {\bibfnamefont
  {M.}~\bibnamefont {Lein}}, \ and\ \bibinfo {author} {\bibfnamefont
  {P.}~\bibnamefont {Lu}},\ }\href {\doibase 10.1103/PhysRevLett.119.033201}
  {\bibfield  {journal} {\bibinfo  {journal} {Phys. Rev. Lett.}\ }\textbf
  {\bibinfo {volume} {119}},\ \bibinfo {pages} {033201} (\bibinfo {year}
  {2017})}\BibitemShut {NoStop}%
\bibitem [{\citenamefont {Cireasa}\ \emph {et~al.}(2015)\citenamefont
  {Cireasa}, \citenamefont {Boguslavskiy}, \citenamefont {Pons}, \citenamefont
  {Wong}, \citenamefont {Descamps}, \citenamefont {Petit}, \citenamefont {Ruf},
  \citenamefont {Thir{\'e}}, \citenamefont {Ferr{\'e}}, \citenamefont {Suarez},
  \citenamefont {Higuet}, \citenamefont {Schmidt}, \citenamefont {Alharbi},
  \citenamefont {L{\'e}gar{\'e}}, \citenamefont {Blanchet}, \citenamefont
  {Fabre}, \citenamefont {Patchkovskii}, \citenamefont {Smirnova},
  \citenamefont {Mairesse},\ and\ \citenamefont {Bhardwaj}}]{Cireasa2015}%
  \BibitemOpen
  \bibfield  {author} {\bibinfo {author} {\bibfnamefont {R.}~\bibnamefont
  {Cireasa}}, \bibinfo {author} {\bibfnamefont {A.~E.}\ \bibnamefont
  {Boguslavskiy}}, \bibinfo {author} {\bibfnamefont {B.}~\bibnamefont {Pons}},
  \bibinfo {author} {\bibfnamefont {M.~C.~H.}\ \bibnamefont {Wong}}, \bibinfo
  {author} {\bibfnamefont {D.}~\bibnamefont {Descamps}}, \bibinfo {author}
  {\bibfnamefont {S.}~\bibnamefont {Petit}}, \bibinfo {author} {\bibfnamefont
  {H.}~\bibnamefont {Ruf}}, \bibinfo {author} {\bibfnamefont {N.}~\bibnamefont
  {Thir{\'e}}}, \bibinfo {author} {\bibfnamefont {A.}~\bibnamefont
  {Ferr{\'e}}}, \bibinfo {author} {\bibfnamefont {J.}~\bibnamefont {Suarez}},
  \bibinfo {author} {\bibfnamefont {J.}~\bibnamefont {Higuet}}, \bibinfo
  {author} {\bibfnamefont {B.~E.}\ \bibnamefont {Schmidt}}, \bibinfo {author}
  {\bibfnamefont {A.~F.}\ \bibnamefont {Alharbi}}, \bibinfo {author}
  {\bibfnamefont {F.}~\bibnamefont {L{\'e}gar{\'e}}}, \bibinfo {author}
  {\bibfnamefont {V.}~\bibnamefont {Blanchet}}, \bibinfo {author}
  {\bibfnamefont {B.}~\bibnamefont {Fabre}}, \bibinfo {author} {\bibfnamefont
  {S.}~\bibnamefont {Patchkovskii}}, \bibinfo {author} {\bibfnamefont
  {O.}~\bibnamefont {Smirnova}}, \bibinfo {author} {\bibfnamefont
  {Y.}~\bibnamefont {Mairesse}}, \ and\ \bibinfo {author} {\bibfnamefont
  {V.~R.}\ \bibnamefont {Bhardwaj}},\ }\href {\doibase 10.1038/nphys3369}
  {\bibfield  {journal} {\bibinfo  {journal} {Nature Physics}\ }\textbf
  {\bibinfo {volume} {11}},\ \bibinfo {pages} {654} (\bibinfo {year}
  {2015})}\BibitemShut {NoStop}%
\bibitem [{\citenamefont {Ayuso}\ \emph {et~al.}(2018)\citenamefont {Ayuso},
  \citenamefont {Decleva}, \citenamefont {Patchkovskii},\ and\ \citenamefont
  {Smirnova}}]{Ayuso_2018}%
  \BibitemOpen
  \bibfield  {author} {\bibinfo {author} {\bibfnamefont {D.}~\bibnamefont
  {Ayuso}}, \bibinfo {author} {\bibfnamefont {P.}~\bibnamefont {Decleva}},
  \bibinfo {author} {\bibfnamefont {S.}~\bibnamefont {Patchkovskii}}, \ and\
  \bibinfo {author} {\bibfnamefont {O.}~\bibnamefont {Smirnova}},\ }\href
  {\doibase 10.1088/1361-6455/aabc95} {\bibfield  {journal} {\bibinfo
  {journal} {Journal of Physics B: Atomic, Molecular and Optical Physics}\
  }\textbf {\bibinfo {volume} {51}},\ \bibinfo {pages} {124002} (\bibinfo
  {year} {2018})}\BibitemShut {NoStop}%
\bibitem [{\citenamefont {Neufeld}\ \emph {et~al.}(2019)\citenamefont
  {Neufeld}, \citenamefont {Ayuso}, \citenamefont {Decleva}, \citenamefont
  {Ivanov}, \citenamefont {Smirnova},\ and\ \citenamefont
  {Cohen}}]{NeufeldPhysRevX}%
  \BibitemOpen
  \bibfield  {author} {\bibinfo {author} {\bibfnamefont {O.}~\bibnamefont
  {Neufeld}}, \bibinfo {author} {\bibfnamefont {D.}~\bibnamefont {Ayuso}},
  \bibinfo {author} {\bibfnamefont {P.}~\bibnamefont {Decleva}}, \bibinfo
  {author} {\bibfnamefont {M.~Y.}\ \bibnamefont {Ivanov}}, \bibinfo {author}
  {\bibfnamefont {O.}~\bibnamefont {Smirnova}}, \ and\ \bibinfo {author}
  {\bibfnamefont {O.}~\bibnamefont {Cohen}},\ }\href {\doibase
  10.1103/PhysRevX.9.031002} {\bibfield  {journal} {\bibinfo  {journal} {Phys.
  Rev. X}\ }\textbf {\bibinfo {volume} {9}},\ \bibinfo {pages} {031002}
  (\bibinfo {year} {2019})}\BibitemShut {NoStop}%
\bibitem [{\citenamefont {Baykusheva}\ \emph {et~al.}(2019)\citenamefont
  {Baykusheva}, \citenamefont {Zindel}, \citenamefont {Svoboda}, \citenamefont
  {Bommeli}, \citenamefont {Ochsner}, \citenamefont {Tehlar},\ and\
  \citenamefont {W{\"o}rner}}]{BaykushevaPNAS}%
  \BibitemOpen
  \bibfield  {author} {\bibinfo {author} {\bibfnamefont {D.}~\bibnamefont
  {Baykusheva}}, \bibinfo {author} {\bibfnamefont {D.}~\bibnamefont {Zindel}},
  \bibinfo {author} {\bibfnamefont {V.}~\bibnamefont {Svoboda}}, \bibinfo
  {author} {\bibfnamefont {E.}~\bibnamefont {Bommeli}}, \bibinfo {author}
  {\bibfnamefont {M.}~\bibnamefont {Ochsner}}, \bibinfo {author} {\bibfnamefont
  {A.}~\bibnamefont {Tehlar}}, \ and\ \bibinfo {author} {\bibfnamefont {H.~J.}\
  \bibnamefont {W{\"o}rner}},\ }\href {\doibase 10.1073/pnas.1907189116}
  {\bibfield  {journal} {\bibinfo  {journal} {Proceedings of the National
  Academy of Sciences}\ }\textbf {\bibinfo {volume} {116}},\ \bibinfo {pages}
  {23923} (\bibinfo {year} {2019})}\BibitemShut {NoStop}%
\bibitem [{\citenamefont {Vampa}\ \emph {et~al.}(2015)\citenamefont {Vampa},
  \citenamefont {Hammond}, \citenamefont {Thir\'e}, \citenamefont {Schmidt},
  \citenamefont {L\'egar\'e}, \citenamefont {McDonald}, \citenamefont {Brabec},
  \citenamefont {Klug},\ and\ \citenamefont {Corkum}}]{VampaPRL}%
  \BibitemOpen
  \bibfield  {author} {\bibinfo {author} {\bibfnamefont {G.}~\bibnamefont
  {Vampa}}, \bibinfo {author} {\bibfnamefont {T.~J.}\ \bibnamefont {Hammond}},
  \bibinfo {author} {\bibfnamefont {N.}~\bibnamefont {Thir\'e}}, \bibinfo
  {author} {\bibfnamefont {B.~E.}\ \bibnamefont {Schmidt}}, \bibinfo {author}
  {\bibfnamefont {F.}~\bibnamefont {L\'egar\'e}}, \bibinfo {author}
  {\bibfnamefont {C.~R.}\ \bibnamefont {McDonald}}, \bibinfo {author}
  {\bibfnamefont {T.}~\bibnamefont {Brabec}}, \bibinfo {author} {\bibfnamefont
  {D.~D.}\ \bibnamefont {Klug}}, \ and\ \bibinfo {author} {\bibfnamefont
  {P.~B.}\ \bibnamefont {Corkum}},\ }\href {\doibase
  10.1103/PhysRevLett.115.193603} {\bibfield  {journal} {\bibinfo  {journal}
  {Phys. Rev. Lett.}\ }\textbf {\bibinfo {volume} {115}},\ \bibinfo {pages}
  {193603} (\bibinfo {year} {2015})}\BibitemShut {NoStop}%
\bibitem [{\citenamefont {Pattanayak}\ \emph {et~al.}(2020)\citenamefont
  {Pattanayak}, \citenamefont {S.},\ and\ \citenamefont {Dixit}}]{AdhipPRA}%
  \BibitemOpen
  \bibfield  {author} {\bibinfo {author} {\bibfnamefont {A.}~\bibnamefont
  {Pattanayak}}, \bibinfo {author} {\bibfnamefont {M.~M.}\ \bibnamefont {S.}},
  \ and\ \bibinfo {author} {\bibfnamefont {G.}~\bibnamefont {Dixit}},\ }\href
  {\doibase 10.1103/PhysRevA.101.013404} {\bibfield  {journal} {\bibinfo
  {journal} {Phys. Rev. A}\ }\textbf {\bibinfo {volume} {101}},\ \bibinfo
  {pages} {013404} (\bibinfo {year} {2020})}\BibitemShut {NoStop}%
\bibitem [{\citenamefont {Mrudul}\ \emph
  {et~al.}(2020{\natexlab{a}})\citenamefont {Mrudul}, \citenamefont
  {Tancogne-Dejean}, \citenamefont {Rubio},\ and\ \citenamefont
  {Dixit}}]{mrudul2020high}%
  \BibitemOpen
  \bibfield  {author} {\bibinfo {author} {\bibfnamefont {M.~S.}\ \bibnamefont
  {Mrudul}}, \bibinfo {author} {\bibfnamefont {N.}~\bibnamefont
  {Tancogne-Dejean}}, \bibinfo {author} {\bibfnamefont {A.}~\bibnamefont
  {Rubio}}, \ and\ \bibinfo {author} {\bibfnamefont {G.}~\bibnamefont
  {Dixit}},\ }\href@noop {} {\bibfield  {journal} {\bibinfo  {journal} {npj
  Computational Materials}\ }\textbf {\bibinfo {volume} {6}},\ \bibinfo {pages}
  {1} (\bibinfo {year} {2020}{\natexlab{a}})}\BibitemShut {NoStop}%
\bibitem [{\citenamefont {Luu}\ \emph {et~al.}(2015)\citenamefont {Luu},
  \citenamefont {Garg}, \citenamefont {Kruchinin}, \citenamefont {Moulet},
  \citenamefont {Hassan},\ and\ \citenamefont
  {Goulielmakis}}]{LuuGoulielmakis2015}%
  \BibitemOpen
  \bibfield  {author} {\bibinfo {author} {\bibfnamefont {T.~T.}\ \bibnamefont
  {Luu}}, \bibinfo {author} {\bibfnamefont {M.}~\bibnamefont {Garg}}, \bibinfo
  {author} {\bibfnamefont {S.~Y.}\ \bibnamefont {Kruchinin}}, \bibinfo {author}
  {\bibfnamefont {A.}~\bibnamefont {Moulet}}, \bibinfo {author} {\bibfnamefont
  {M.~T.}\ \bibnamefont {Hassan}}, \ and\ \bibinfo {author} {\bibfnamefont
  {E.}~\bibnamefont {Goulielmakis}},\ }\href {\doibase 10.1038/nature14456}
  {\bibfield  {journal} {\bibinfo  {journal} {Nature}\ }\textbf {\bibinfo
  {volume} {521}},\ \bibinfo {pages} {498} (\bibinfo {year}
  {2015})}\BibitemShut {NoStop}%
\bibitem [{\citenamefont {Langer}\ \emph {et~al.}(2018)\citenamefont {Langer},
  \citenamefont {Schmid}, \citenamefont {Schlauderer}, \citenamefont {Gmitra},
  \citenamefont {Fabian}, \citenamefont {Nagler}, \citenamefont {Sch{\"u}ller},
  \citenamefont {Korn}, \citenamefont {Hawkins}, \citenamefont {Steiner},
  \citenamefont {Huttner}, \citenamefont {Koch}, \citenamefont {Kira},\ and\
  \citenamefont {Huber}}]{LangerHuber2018}%
  \BibitemOpen
  \bibfield  {author} {\bibinfo {author} {\bibfnamefont {F.}~\bibnamefont
  {Langer}}, \bibinfo {author} {\bibfnamefont {C.~P.}\ \bibnamefont {Schmid}},
  \bibinfo {author} {\bibfnamefont {S.}~\bibnamefont {Schlauderer}}, \bibinfo
  {author} {\bibfnamefont {M.}~\bibnamefont {Gmitra}}, \bibinfo {author}
  {\bibfnamefont {J.}~\bibnamefont {Fabian}}, \bibinfo {author} {\bibfnamefont
  {P.}~\bibnamefont {Nagler}}, \bibinfo {author} {\bibfnamefont
  {C.}~\bibnamefont {Sch{\"u}ller}}, \bibinfo {author} {\bibfnamefont
  {T.}~\bibnamefont {Korn}}, \bibinfo {author} {\bibfnamefont {P.~G.}\
  \bibnamefont {Hawkins}}, \bibinfo {author} {\bibfnamefont {J.~T.}\
  \bibnamefont {Steiner}}, \bibinfo {author} {\bibfnamefont {U.}~\bibnamefont
  {Huttner}}, \bibinfo {author} {\bibfnamefont {S.~W.}\ \bibnamefont {Koch}},
  \bibinfo {author} {\bibfnamefont {M.}~\bibnamefont {Kira}}, \ and\ \bibinfo
  {author} {\bibfnamefont {R.}~\bibnamefont {Huber}},\ }\href {\doibase
  10.1038/s41586-018-0013-6} {\bibfield  {journal} {\bibinfo  {journal}
  {Nature}\ }\textbf {\bibinfo {volume} {557}},\ \bibinfo {pages} {76}
  (\bibinfo {year} {2018})}\BibitemShut {NoStop}%
\bibitem [{\citenamefont {Jimenez-Galan}\ \emph {et~al.}(2020)\citenamefont
  {Jimenez-Galan}, \citenamefont {Silva}, \citenamefont {Smirnova},\ and\
  \citenamefont {Ivanov}}]{Alvaro2020subcycle}%
  \BibitemOpen
  \bibfield  {author} {\bibinfo {author} {\bibfnamefont {A.}~\bibnamefont
  {Jimenez-Galan}}, \bibinfo {author} {\bibfnamefont {R.}~\bibnamefont
  {Silva}}, \bibinfo {author} {\bibfnamefont {O.}~\bibnamefont {Smirnova}}, \
  and\ \bibinfo {author} {\bibfnamefont {M.}~\bibnamefont {Ivanov}},\
  }\href@noop {} {\enquote {\bibinfo {title} {Sub-cycle valleytronics: control
  of valley polarization using few-cycle linearly polarized pulses},}\ }
  (\bibinfo {year} {2020}),\ \Eprint {http://arxiv.org/abs/2005.10196}
  {arXiv:2005.10196 [physics.optics]} \BibitemShut {NoStop}%
\bibitem [{\citenamefont {Mrudul}\ \emph
  {et~al.}(2020{\natexlab{b}})\citenamefont {Mrudul}, \citenamefont
  {Jim{\'e}nez-Gal{\'a}n}, \citenamefont {Ivanov},\ and\ \citenamefont
  {Dixit}}]{jimenez2020light}%
  \BibitemOpen
  \bibfield  {author} {\bibinfo {author} {\bibfnamefont {M.~S.}\ \bibnamefont
  {Mrudul}}, \bibinfo {author} {\bibfnamefont {{\'A}.}~\bibnamefont
  {Jim{\'e}nez-Gal{\'a}n}}, \bibinfo {author} {\bibfnamefont {M.}~\bibnamefont
  {Ivanov}}, \ and\ \bibinfo {author} {\bibfnamefont {G.}~\bibnamefont
  {Dixit}},\ }\href@noop {} {\bibfield  {journal} {\bibinfo  {journal} {arXiv
  preprint arXiv:2011.04973}\ } (\bibinfo {year}
  {2020}{\natexlab{b}})}\BibitemShut {NoStop}%
\bibitem [{\citenamefont {Uzan}\ \emph {et~al.}(2020)\citenamefont {Uzan},
  \citenamefont {Orenstein}, \citenamefont {Jim{\'e}nez-Gal{\'a}n},
  \citenamefont {McDonald}, \citenamefont {Silva}, \citenamefont {Bruner},
  \citenamefont {Klimkin}, \citenamefont {Blanchet}, \citenamefont
  {Arusi-Parpar}, \citenamefont {Kr{\"u}ger}, \citenamefont {Rubtsov},
  \citenamefont {Smirnova}, \citenamefont {Ivanov}, \citenamefont {Yan},
  \citenamefont {Brabec},\ and\ \citenamefont {Dudovich}}]{Uzan2020}%
  \BibitemOpen
  \bibfield  {author} {\bibinfo {author} {\bibfnamefont {A.~J.}\ \bibnamefont
  {Uzan}}, \bibinfo {author} {\bibfnamefont {G.}~\bibnamefont {Orenstein}},
  \bibinfo {author} {\bibfnamefont {{\'A}.}~\bibnamefont
  {Jim{\'e}nez-Gal{\'a}n}}, \bibinfo {author} {\bibfnamefont {C.}~\bibnamefont
  {McDonald}}, \bibinfo {author} {\bibfnamefont {R.~E.~F.}\ \bibnamefont
  {Silva}}, \bibinfo {author} {\bibfnamefont {B.~D.}\ \bibnamefont {Bruner}},
  \bibinfo {author} {\bibfnamefont {N.~D.}\ \bibnamefont {Klimkin}}, \bibinfo
  {author} {\bibfnamefont {V.}~\bibnamefont {Blanchet}}, \bibinfo {author}
  {\bibfnamefont {T.}~\bibnamefont {Arusi-Parpar}}, \bibinfo {author}
  {\bibfnamefont {M.}~\bibnamefont {Kr{\"u}ger}}, \bibinfo {author}
  {\bibfnamefont {A.~N.}\ \bibnamefont {Rubtsov}}, \bibinfo {author}
  {\bibfnamefont {O.}~\bibnamefont {Smirnova}}, \bibinfo {author}
  {\bibfnamefont {M.}~\bibnamefont {Ivanov}}, \bibinfo {author} {\bibfnamefont
  {B.}~\bibnamefont {Yan}}, \bibinfo {author} {\bibfnamefont {T.}~\bibnamefont
  {Brabec}}, \ and\ \bibinfo {author} {\bibfnamefont {N.}~\bibnamefont
  {Dudovich}},\ }\href {\doibase 10.1038/s41566-019-0574-4} {\bibfield
  {journal} {\bibinfo  {journal} {Nature Photonics}\ }\textbf {\bibinfo
  {volume} {14}},\ \bibinfo {pages} {183} (\bibinfo {year} {2020})}\BibitemShut
  {NoStop}%
\bibitem [{\citenamefont {Lakhotia}\ \emph {et~al.}(2020)\citenamefont
  {Lakhotia}, \citenamefont {Kim}, \citenamefont {Zhan}, \citenamefont {Hu},
  \citenamefont {Meng},\ and\ \citenamefont
  {Goulielmakis}}]{LakhotiaGoulielmakis2020}%
  \BibitemOpen
  \bibfield  {author} {\bibinfo {author} {\bibfnamefont {H.}~\bibnamefont
  {Lakhotia}}, \bibinfo {author} {\bibfnamefont {H.~Y.}\ \bibnamefont {Kim}},
  \bibinfo {author} {\bibfnamefont {M.}~\bibnamefont {Zhan}}, \bibinfo {author}
  {\bibfnamefont {S.}~\bibnamefont {Hu}}, \bibinfo {author} {\bibfnamefont
  {S.}~\bibnamefont {Meng}}, \ and\ \bibinfo {author} {\bibfnamefont
  {E.}~\bibnamefont {Goulielmakis}},\ }\href {\doibase
  10.1038/s41586-020-2429-z} {\bibfield  {journal} {\bibinfo  {journal}
  {Nature}\ }\textbf {\bibinfo {volume} {583}},\ \bibinfo {pages} {55}
  (\bibinfo {year} {2020})}\BibitemShut {NoStop}%
\bibitem [{\citenamefont {Mrudul}\ \emph {et~al.}(2019)\citenamefont {Mrudul},
  \citenamefont {Pattanayak}, \citenamefont {Ivanov},\ and\ \citenamefont
  {Dixit}}]{pattanayak2019direct}%
  \BibitemOpen
  \bibfield  {author} {\bibinfo {author} {\bibfnamefont {M.~S.}\ \bibnamefont
  {Mrudul}}, \bibinfo {author} {\bibfnamefont {A.}~\bibnamefont {Pattanayak}},
  \bibinfo {author} {\bibfnamefont {M.}~\bibnamefont {Ivanov}}, \ and\ \bibinfo
  {author} {\bibfnamefont {G.}~\bibnamefont {Dixit}},\ }\href@noop {}
  {\bibfield  {journal} {\bibinfo  {journal} {Physical Review A}\ }\textbf
  {\bibinfo {volume} {100}},\ \bibinfo {pages} {043420} (\bibinfo {year}
  {2019})}\BibitemShut {NoStop}%
\bibitem [{\citenamefont {Silva}\ \emph {et~al.}(2018)\citenamefont {Silva},
  \citenamefont {Blinov}, \citenamefont {Rubtsov}, \citenamefont {Smirnova},\
  and\ \citenamefont {Ivanov}}]{Silva2018}%
  \BibitemOpen
  \bibfield  {author} {\bibinfo {author} {\bibfnamefont {R.~E.~F.}\
  \bibnamefont {Silva}}, \bibinfo {author} {\bibfnamefont {I.~V.}\ \bibnamefont
  {Blinov}}, \bibinfo {author} {\bibfnamefont {A.~N.}\ \bibnamefont {Rubtsov}},
  \bibinfo {author} {\bibfnamefont {O.}~\bibnamefont {Smirnova}}, \ and\
  \bibinfo {author} {\bibfnamefont {M.}~\bibnamefont {Ivanov}},\ }\href
  {\doibase 10.1038/s41566-018-0129-0} {\bibfield  {journal} {\bibinfo
  {journal} {Nature Photonics}\ }\textbf {\bibinfo {volume} {12}},\ \bibinfo
  {pages} {266} (\bibinfo {year} {2018})}\BibitemShut {NoStop}%
\bibitem [{\citenamefont {Bauer}\ and\ \citenamefont
  {Hansen}(2018)}]{BauerPRLtopoedge}%
  \BibitemOpen
  \bibfield  {author} {\bibinfo {author} {\bibfnamefont {D.}~\bibnamefont
  {Bauer}}\ and\ \bibinfo {author} {\bibfnamefont {K.~K.}\ \bibnamefont
  {Hansen}},\ }\href {\doibase 10.1103/PhysRevLett.120.177401} {\bibfield
  {journal} {\bibinfo  {journal} {Phys. Rev. Lett.}\ }\textbf {\bibinfo
  {volume} {120}},\ \bibinfo {pages} {177401} (\bibinfo {year}
  {2018})}\BibitemShut {NoStop}%
\bibitem [{\citenamefont {Silva}\ \emph {et~al.}(2019)\citenamefont {Silva},
  \citenamefont {Jim{\'e}nez-Gal{\'a}n}, \citenamefont {Amorim}, \citenamefont
  {Smirnova},\ and\ \citenamefont {Ivanov}}]{Silva2019}%
  \BibitemOpen
  \bibfield  {author} {\bibinfo {author} {\bibfnamefont {R.~E.~F.}\
  \bibnamefont {Silva}}, \bibinfo {author} {\bibfnamefont {{\'A}.}~\bibnamefont
  {Jim{\'e}nez-Gal{\'a}n}}, \bibinfo {author} {\bibfnamefont {B.}~\bibnamefont
  {Amorim}}, \bibinfo {author} {\bibfnamefont {O.}~\bibnamefont {Smirnova}}, \
  and\ \bibinfo {author} {\bibfnamefont {M.}~\bibnamefont {Ivanov}},\ }\href
  {\doibase 10.1038/s41566-019-0516-1} {\bibfield  {journal} {\bibinfo
  {journal} {Nature Photonics}\ }\textbf {\bibinfo {volume} {13}},\ \bibinfo
  {pages} {849} (\bibinfo {year} {2019})}\BibitemShut {NoStop}%
\bibitem [{\citenamefont {Aubry}\ and\ \citenamefont {Andre}(1980)}]{Aubry}%
  \BibitemOpen
  \bibfield  {author} {\bibinfo {author} {\bibfnamefont {S.}~\bibnamefont
  {Aubry}}\ and\ \bibinfo {author} {\bibfnamefont {G.}~\bibnamefont {Andre}},\
  }\href@noop {} {\bibfield  {journal} {\bibinfo  {journal} {Ann. Israel Phys.
  Soc}\ }\textbf {\bibinfo {volume} {3}},\ \bibinfo {pages} {18} (\bibinfo
  {year} {1980})}\BibitemShut {NoStop}%
\bibitem [{\citenamefont {Harper}(1955)}]{Harper_1955}%
  \BibitemOpen
  \bibfield  {author} {\bibinfo {author} {\bibfnamefont {P.~G.}\ \bibnamefont
  {Harper}},\ }\href {\doibase 10.1088/0370-1298/68/10/304} {\bibfield
  {journal} {\bibinfo  {journal} {Proceedings of the Physical Society. Section
  A}\ }\textbf {\bibinfo {volume} {68}},\ \bibinfo {pages} {874} (\bibinfo
  {year} {1955})}\BibitemShut {NoStop}%
\bibitem [{\citenamefont {Sanchez-Palencia}\ and\ \citenamefont
  {Lewenstein}(2010)}]{Sanchez-Palencia2010}%
  \BibitemOpen
  \bibfield  {author} {\bibinfo {author} {\bibfnamefont {L.}~\bibnamefont
  {Sanchez-Palencia}}\ and\ \bibinfo {author} {\bibfnamefont {M.}~\bibnamefont
  {Lewenstein}},\ }\href {\doibase 10.1038/nphys1507} {\bibfield  {journal}
  {\bibinfo  {journal} {Nature Physics}\ }\textbf {\bibinfo {volume} {6}},\
  \bibinfo {pages} {87} (\bibinfo {year} {2010})}\BibitemShut {NoStop}%
\bibitem [{\citenamefont {Roati}\ \emph {et~al.}(2008)\citenamefont {Roati},
  \citenamefont {D'Errico}, \citenamefont {Fallani}, \citenamefont {Fattori},
  \citenamefont {Fort}, \citenamefont {Zaccanti}, \citenamefont {Modugno},
  \citenamefont {Modugno},\ and\ \citenamefont {Inguscio}}]{Roati2008}%
  \BibitemOpen
  \bibfield  {author} {\bibinfo {author} {\bibfnamefont {G.}~\bibnamefont
  {Roati}}, \bibinfo {author} {\bibfnamefont {C.}~\bibnamefont {D'Errico}},
  \bibinfo {author} {\bibfnamefont {L.}~\bibnamefont {Fallani}}, \bibinfo
  {author} {\bibfnamefont {M.}~\bibnamefont {Fattori}}, \bibinfo {author}
  {\bibfnamefont {C.}~\bibnamefont {Fort}}, \bibinfo {author} {\bibfnamefont
  {M.}~\bibnamefont {Zaccanti}}, \bibinfo {author} {\bibfnamefont
  {G.}~\bibnamefont {Modugno}}, \bibinfo {author} {\bibfnamefont
  {M.}~\bibnamefont {Modugno}}, \ and\ \bibinfo {author} {\bibfnamefont
  {M.}~\bibnamefont {Inguscio}},\ }\href {\doibase 10.1038/nature07071}
  {\bibfield  {journal} {\bibinfo  {journal} {Nature}\ }\textbf {\bibinfo
  {volume} {453}},\ \bibinfo {pages} {895} (\bibinfo {year}
  {2008})}\BibitemShut {NoStop}%
\bibitem [{\citenamefont {Ivanov}\ \emph {et~al.}(1996)\citenamefont {Ivanov},
  \citenamefont {Seideman}, \citenamefont {Corkum}, \citenamefont {Ilkov},\
  and\ \citenamefont {Dietrich}}]{MishaPRA1996}%
  \BibitemOpen
  \bibfield  {author} {\bibinfo {author} {\bibfnamefont {M.}~\bibnamefont
  {Ivanov}}, \bibinfo {author} {\bibfnamefont {T.}~\bibnamefont {Seideman}},
  \bibinfo {author} {\bibfnamefont {P.}~\bibnamefont {Corkum}}, \bibinfo
  {author} {\bibfnamefont {F.}~\bibnamefont {Ilkov}}, \ and\ \bibinfo {author}
  {\bibfnamefont {P.}~\bibnamefont {Dietrich}},\ }\href {\doibase
  10.1103/PhysRevA.54.1541} {\bibfield  {journal} {\bibinfo  {journal} {Phys.
  Rev. A}\ }\textbf {\bibinfo {volume} {54}},\ \bibinfo {pages} {1541}
  (\bibinfo {year} {1996})}\BibitemShut {NoStop}%
\bibitem [{\citenamefont {Dietrich}\ \emph {et~al.}(1996)\citenamefont
  {Dietrich}, \citenamefont {Ivanov}, \citenamefont {Ilkov},\ and\
  \citenamefont {Corkum}}]{DietrichIvanovPRL1996}%
  \BibitemOpen
  \bibfield  {author} {\bibinfo {author} {\bibfnamefont {P.}~\bibnamefont
  {Dietrich}}, \bibinfo {author} {\bibfnamefont {M.~Y.}\ \bibnamefont
  {Ivanov}}, \bibinfo {author} {\bibfnamefont {F.~A.}\ \bibnamefont {Ilkov}}, \
  and\ \bibinfo {author} {\bibfnamefont {P.~B.}\ \bibnamefont {Corkum}},\
  }\href {\doibase 10.1103/PhysRevLett.77.4150} {\bibfield  {journal} {\bibinfo
   {journal} {Phys. Rev. Lett.}\ }\textbf {\bibinfo {volume} {77}},\ \bibinfo
  {pages} {4150} (\bibinfo {year} {1996})}\BibitemShut {NoStop}%
\bibitem [{\citenamefont {Wu}\ \emph {et~al.}(2015)\citenamefont {Wu},
  \citenamefont {Ghimire}, \citenamefont {Reis}, \citenamefont {Schafer},\ and\
  \citenamefont {Gaarde}}]{MWuPRA}%
  \BibitemOpen
  \bibfield  {author} {\bibinfo {author} {\bibfnamefont {M.}~\bibnamefont
  {Wu}}, \bibinfo {author} {\bibfnamefont {S.}~\bibnamefont {Ghimire}},
  \bibinfo {author} {\bibfnamefont {D.~A.}\ \bibnamefont {Reis}}, \bibinfo
  {author} {\bibfnamefont {K.~J.}\ \bibnamefont {Schafer}}, \ and\ \bibinfo
  {author} {\bibfnamefont {M.~B.}\ \bibnamefont {Gaarde}},\ }\href {\doibase
  10.1103/PhysRevA.91.043839} {\bibfield  {journal} {\bibinfo  {journal} {Phys.
  Rev. A}\ }\textbf {\bibinfo {volume} {91}},\ \bibinfo {pages} {043839}
  (\bibinfo {year} {2015})}\BibitemShut {NoStop}%
\end{thebibliography}
%

\end{document}